\documentclass[aps,prd,twocolumn,reprint,superscriptaddress,nofootinbib,floatfix]{revtex4-1}

\usepackage{amsmath}
\usepackage{bm}
\usepackage{amsfonts}
\usepackage{graphicx}
\usepackage{epstopdf}
\usepackage{array}
\usepackage{soul}
\usepackage{color}
\usepackage{hyperref}
\usepackage{amssymb}
\usepackage{comment}
\usepackage{subcaption}
\usepackage{float}
\usepackage{silence}
\usepackage{caption}
\DeclareMathOperator\erfc{erfc}

\smallskipamount

\enlargethispage{\baselineskip}

\hypersetup{
    colorlinks=true,
    urlcolor=blue,
    citecolor=blue,
    linkcolor=blue,
    menucolor=blue,
    anchorcolor=blue,
    filecolor=blue
}

\everymath{\displaystyle}
\begin{document}

\title{Constraining the primordial black hole abundance through Big-Bang nucleosynthesis}

\newcommand{\UniNA}{\affiliation{Universit\`a degli Studi di Napoli ``Federico II'',
                Via Cintia 26, 80126 Napoli, Italy}}
\newcommand{\INFN}{\affiliation{INFN sezione di Napoli, Via Cintia 1, 80126 Napoli, Italy}}
\newcommand{\SSM}{\affiliation{Scuola Superiore Meridionale, Largo S.\ Marcellino 10, I-80138 Napoli, Italy}}
\newcommand{\TDLI}{\affiliation{Tsung-Dao Lee Institute (TDLI), No.\ 1 Lisuo Road, 201210 Shanghai, China}}
\newcommand{\SJTU}{\affiliation{School of Physics and Astronomy, Shanghai Jiao Tong University, \\ Dongchuan Road 800, 201240 Shanghai, China}}

\author{Andrea Boccia}
\email{andrea.boccia-ssm@unina.it}
\SSM \UniNA

\author{Fabio Iocco}
\email{fabio.iocco.unina@gmail.com}
\UniNA \INFN

\author{Luca Visinelli}
\email{luca.visinelli@sjtu.edu.cn}
\TDLI \SJTU

\date{\today}

\begin{abstract}
We investigate the scenario in which primordial black holes (PBHs) with masses $M_{\rm PBH} \lesssim 10^9$\,g undergo Hawking evaporation, around the Big-Bang nucleosynthesis (BBN) epoch. The evaporation process modifies the Universe's expansion rate and the baryon-to-photon ratio, leading to an alteration of the primordial abundance of light nuclei. 
We present numerical solutions for the set of equations describing this physics, considering different values of PBH masses and abundances at their formation, showing how their evaporation impacts the abundances of light nuclei, obtained by incorporating the non-standard Hubble rate and baryon-to-photon ratio into the BBN code \texttt{PArthENoPE}. 
The results are then used to place upper bounds 
for the PBH relative abundance at formation in the range $10^8{\rm\,g} \lesssim M_{\rm PBH} \lesssim 10^9$\,g, providing the strongest constraints existing to-date in this mass range.

\end{abstract}

\maketitle

\section{Introduction}

Primordial black holes (PBHs) are hypothetical objects formed in the very early Universe via gravitational collapse of large density fluctuations in the hot plasma~\cite{Zeldovich:1967lct, Hawking:1971ei, Carr:1974nx}. This occurs when an overdense region with an amplitude greater than a critical value
%$(\delta_c \gtrsim w)$
re-enters the horizon~\cite{Carr:1975qj, Shibata:1999zs, Niemeyer:1999ak, Musco:2004ak}. These large inhomogeneities of order unity, produced through different mechanisms, allow for the formation of PBHs with a mass of the same order as the total horizon mass at that time. For this, more massive PBHs form at later times~\cite{Carr:2020erq}.

Owing to the mechanism described above, PBHs can in principle cover a wide mass spectrum, being suitable both as dark matter (DM) candidates~\cite{Carr:2016drx, Green:2020jor} and as seeds for supermassive black holes~\cite{Duechting:2004dk}. This last statement, however, applies only to those PBHs which are stable over cosmological times, as it is commonly thought that black holes should emit Hawking radiation and eventually evaporate due to quantum effects~\cite{Hawking:1975vcx}. In this framework, it is expected that PBHs lighter than $M_{\rm PBH} \lesssim 10^{15}$\,g would have evaporated by now. Even though these light PBHs cannot solve the late-time cosmology puzzles, it is still worth studying them as they could have had an important role in the thermal history of the Universe, leaving signatures of their existence onto early Universe-related cosmological observables.

For instance, the energy density of light PBHs formed very early on would dilute at a lower rate than the surrounding relativistic plasma as the Universe expands, triggering an epoch of early matter domination that would last down to PBH evaporation time $t_{\rm evap}$. 
The condition for this to happen is that the energy density of PBHs overcomes that of radiation before $t_{\rm evap}$, condition which can be expressed in terms of the abundance at formation of PBHs $\beta$ and the temperature of the Universe at their formation and evaporation, respectively $T_f$ and $T_{\rm evap}$. An early matter domination is achieved if $\beta$ exceeds the critical value $\beta_{\rm crit}\equiv T_{\rm evap}/T_f$.

Here, we consider a different aspect, namely the consequences of an expansion rate and baryon--to--photon ratio, altered  due to the presence of a primordial fraction of PBHs. Even for values of the PBH abundance at formation below the threshold value $\beta \leq \beta_{\rm crit}$, PBHs could have had an effect on the expansion rate at the onset of nuclear reactions. In addition, the spectrum of highly relativistic particles injected by the PBH evaporation into the cosmological plasma also alters its entropy, thus modifying the baryon-to-photon ratio.
These two physical effects combine inducing an alteration of the reaction rates, thus potentially modifying the primordial abundance of light nuclei yields. 
We exploit this physics to derive new bounds on the PBH parameter space $(\beta, M_{\rm PBH})$. %, which have been overlooked in previous literature.

In this work, we focus on a particular class of light PBHs of mass $M_{\rm PBH} \lesssim 10^{9}$\,g that would have evaporated by the time of Big-Bang nucleosynthesis (BBN). 
The PBHs we study therefore evaporate when the Universe plasma temperature is $40 \, \rm MeV \gtrsim T \gtrsim 1 \, \rm MeV$. In this temperature range the products of evaporation rapidly interact with the environment, fully thermalizing by the time BBN proper starts, thus enabling us to study primordial nucleosynthesis without the inclusion of non--thermal effects, which has been addressed in other works for different physical PBH conditions ~\cite{Kawasaki:2004qu, Kawasaki:2017bqm, Kawasaki:2020qxm}.
It is also important to stress here that the PBH abundance we explore are such that they never come to dominate the energy density of the Universe, PBH energy density at most reaching the Universe's ``{\tt standard}'' radiation density. The numerical resolution of the set of coupled BBN equations is provided by employing the latest version (v3.0) of the code Public Algorithm Evaluating the Nucleosynthesis of Primordial Elements or \texttt{PArthENoPE}~\cite{Pisanti:2007hk, Consiglio:2017pot, 2022CoPhC.27108205G}.\footnote{The code is publicly available at \href{http://parthenope.na.infn.it}{http://parthenope.na.infn.it}.} While we do not modify the BBN reactions described in \texttt{PArthENoPE}, we input the background evolution of the Hubble rate and the value of the baryon-to-photon ratio at $T=10$\,MeV, here $\eta_i$, into the code to account for the modified dynamics in the presence of a PBH population.

This paper is organized as follows. In Sec.~\ref{sec:methods} we discuss the basic equations and the methods used throughout the analysis. Results are presented in Sec.~\ref{sec:results} and discussed in Sec.~\ref{sec:discussion}. Finally, conclusions are drawn in Sec.~\ref{sec:conclusions}. We work with units $\hbar = c = k_B =1$.

\section{Methods}
\label{sec:methods}

\subsection{Background cosmology}

We consider an expanding Universe described by a flat Friedmann-Lema{\^i}tre-Robertson-Walker metric of scale factor $a = a(t)$, where the cosmic time $t$ enters the definition of the Hubble rate $H \equiv \dot a/a$ as a dot over any time-dependent quantity. The Friedmann equation relating the Hubble rate with the total energy content $\rho_{\rm tot}$ at time $t$ is $3H^2 = 8\pi G\rho_{\rm tot}$. In the standard cosmological scenario, the evolution of the Universe in its early stages is dominated by the energy density of the relativistic component,
\begin{equation}
    \label{eq:rho_r}
    \rho_r (T) = \frac{\pi^2}{30} g_*(T)T^4\,,
\end{equation}
with $g_*(T)$ being the number of relativistic degrees of freedom at temperature $T$. As the Universe expands and the scale factor $a$ increases, the relativistic component cools off due to the relation $T \propto 1/a$ that holds for the adiabatic regime.

The model described so far through Eq.~\eqref{eq:rho_r} does not include the contribution from non-relativistic matter, which becomes relevant at around matter-radiation equality (``eq'') characterized by the redshift $1 + z_{\rm eq} \equiv \Omega_m/\Omega_r \sim 3400$. Here, $\Omega_m$ and $\Omega_r$ are the present abundances of matter and radiation, respectively. For redshifts $z < z_{\rm eq}$, pressure-less matter overcomes the share in radiation in governing the Hubble rate.

The picture above potentially modifies once a population of PBHs is included, adding up to non-relativistic matter component that is present since the earliest stages. The PBH fractional abundance at formation with respect to the total energy density is
\begin{equation}
    \label{beta}
    \beta \equiv \frac{\rho_{\rm PBH}(t_f)}{\rho_{\rm tot}(T_f)} \simeq \frac{\rho_{\rm PBH}(t_f)}{\rho_r(T_f)}\,,
\end{equation}
where $T_f$ is the temperature corresponding to time $t_f$ and the second equality holds for a radiation-dominated cosmology. Assuming that the PBHs have a monochromatic mass distribution centered at the value $M_{\rm PBH}$, their energy density at any time $t > t_f$ is found as $\rho_{\rm PBH}(t) = M_{\rm PBH}n_{\rm PBH}(t)$, where $n_{\rm PBH}(t)$ is the number density. Moreover, if the primordial power spectrum has a Gaussian shape, the initial PBH abundance is
\begin{equation}
    \beta \sim \erfc \left[ \frac{\delta_c}{\sqrt{2} \delta_{\rm rms}}\right]\,,
\end{equation}
where $\delta_{\rm rms}$ the root mean square amplitude of the primordial power spectrum. The Hubble rate at formation is related to the PBH mass as~\cite{Carr:2020erq}
\begin{equation}
\label{eq:Hf}
    H_f = \gamma/(2 GM_{\rm PBH})\,,    
\end{equation}
when the horizon mass collapses into the PBH by a fraction $\gamma$, which is $\gamma \sim 0.2$ for a radiation-dominated scenario~\cite{Carr:1975qj}. The PBH energy density at formation is then
\begin{equation}
    \rho_{\rm PBH}(t_f) = \frac{3 \beta \gamma^2}{32 \pi \,G^3\,M_{\rm PBH}^2}\,.
\end{equation}
Given the uncertainty in $\gamma$ and the number of degrees of freedom at PBH formation, $g_{*f}$, the rescaled abundance is introduced as
\begin{equation}
    \label{eq:betaprime}
    \beta' \equiv \gamma^{1/2}\,\left(\frac{g_{*f}}{106.75}\right)^{-1/4}\,\beta\,,
\end{equation}
so that the results in previous literature can be compared~\cite{Carr:2009jm, Carr:2020gox, Luo:2020dlg}. Strictly speaking, the scaling in Eq.~\eqref{eq:betaprime} allows to express a combination of $\beta$ and $\gamma$ as a unique parameter only for models of PBHs that have not yet evaporated, and it is adopted for models with lighter black holes to draw a comparison of the constraints. For this, we also adopt it for this work, setting the reduced Hubble constant $h = 0.674$. Nevertheless, our results explicitly depend on the choice of the parameters $\beta$ and $\gamma$ which enter the initial condition for solving the coupled Boltzmann equation, as discussed below. If the cosmic expansion after PBH formation is adiabatic, the PBH number density scales with the entropy density and it is given by
\begin{equation}
    \label{eq:nPBH}
    n_{\rm PBH}(t) = \beta\,\frac{T_f}{T}\,\frac{\rho_r(T)}{M_{\rm PBH}}\,.
\end{equation}
In this scenario, the PBH component modifies the Hubble expansion rate and might even trigger an early matter-dominated epoch if the energy density in PBHs prevails over that in radiation. Using Eq.~\eqref{eq:nPBH}, this occurs after the hypothetical early matter-radiation equality (``Eeq''), corresponding to temperatures below $T \lesssim T_{\rm Eeq} \equiv \beta T_f$.

An early matter-dominated period is avoided if the evaporation of PBHs occurs prior the hypothetical early matter-radiation equality. The spectrum of particles radiated by an evaporating PBH peaks at the Hawking temperature~\cite{Hawking:1975vcx}
\begin{equation}
    T_H = \frac{1}{8 \pi G M_{\rm PBH}}\,.
\end{equation}
For a non rotating and charge-less PBH~\cite{Mirbabayi:2019uph, DeLuca:2019buf}, the rate of evaporation is~\cite{Page:1976wx}
\begin{equation}
    \label{eq:Mdot}
    \dot{M}_{\rm PBH} = - \frac{\mathcal{G}\,g_H}{30720 \pi G^2 M_{\rm PBH}^2}\,.
\end{equation}
Here, $\mathcal{G}\sim 3.8 $ is a graybody factor that accounts for deviations from a black body spectrum due to the scattering of the emitted radiation on the gravitational field of the PBH, and $g_H = g_H(T_H)$ is the spin-weighted number of degrees of freedom at temperature $T_H$. Summing all contributions from SM particles gives $g_H \simeq 102.6$~\cite{Mazde:2022sdx}, including the effect of mutual repulsion between charged particles~\cite{Page:1977um}. The corresponding evaporation time obtained from integrating Eq.~\eqref{eq:Mdot} is
\begin{equation}
    \label{eq:tevap}
    t_{\rm evap} = 10240 \pi G^2 M_{\rm PBH}^3/(\mathcal{G}\, g_H)\,.
\end{equation}
The condition over which an early matter-dominated epoch is achieved for some period in terms of $\beta$ reads
\begin{equation}
    \label{eq:betacrit}
    \beta \gtrsim \beta_{\rm crit} \equiv T_{\rm evap}/T_f\,,
\end{equation}
with $T_{\rm evap}$ being the temperature of the plasma at PBH evaporation. 
In the parameter space we consider, a matter-dominated era never occurs, the Universe always being radiation dominated, although the actual amount of radiation present is altered with respect to the standard case by the PBH evaporation. 

The time evolution of the energy densities in radiation and PBHs is described by a set of coupled Boltzmann equations in an expanding Universe. For the PBH mass range considered here, the Hawking temperature is $T_H \sim \mathcal{O}(10 \, \rm{TeV})$ so that all SM products of the evaporation are in the form of radiation. For this, the continuity equations for $\rho_r$ and $\rho_{\rm PBH}$ have the bulk expressions
\begin{eqnarray}
    \label{contbh}
    \dot{\rho}_{\rm PBH} + 3H \rho_{\rm PBH}  &=& \frac{\dot{M}_{\rm PBH}}{M_{\rm PBH}}\, \rho_{\rm PBH}\,,\\
    \label{contr}
    \dot{\rho}_r + 4H \rho_r &=& - \frac{\dot{M}_{\rm PBH}}{M_{\rm PBH}}\, \rho_{\rm PBH}\,.
\end{eqnarray}
Here, the term proportional to the Hubble rate gives the cosmic dilution, while the mass-loss rate is given in Eq.~\eqref{eq:Mdot}. Both the energy densities contribute to the Hubble rate as
\begin{equation}
    \label{hub}
    H^2 = \frac{8 \pi G}{3}\left(\rho_r + \rho_{\rm PBH}\right)\,.
\end{equation}
For a given temperature above $T_{\rm evap}$, the expansion rate in the modified history is then higher than in the standard scenario because of the presence of the PBH component. Combining Eqs.~\eqref{contbh} and~\eqref{contr} with the Hubble rate in Eq.~\eqref{hub}, gives the evolution of the PBH density,
\begin{equation}
    \label{eq:PBHabundance}
    \frac{{\rm d} \rho_{\rm PBH}}{{\rm d}T} - \frac{3}{T}\frac{H}{H_{\rm eff}} \rho_{\rm PBH} = - \frac{\dot{M}_{\rm PBH}}{M_{\rm PBH}} \frac{\rho_{\rm PBH}}{T H_{\rm eff}}\,,
\end{equation}
where the effective Hubble rate is
\begin{equation}
    H_{\rm eff} = H + \frac{\dot{M}_{\rm PBH}}{M_{\rm PBH}} \frac{\rho_{\rm PBH}}{4 \rho_r}\,.
\end{equation}
Eq.~\eqref{eq:PBHabundance} accounts for the evaporation of the PBH population and the energy injection into the radiation component through the function $H_{\rm eff}$. When considering the effects of the black holes onto the expansion rate, we solve the equation with the initial condition in Eq.~\eqref{beta}. One key expression related to Eq.~\eqref{contr} and used in the derivations below concerns the evolution of the entropy density $s(T)$, which is here given as
\begin{equation}
    \label{eq:sT}
    \dot{s} + 3H_{\rm eff}\,s = 0\,.
\end{equation}

\subsection{Abundances of light nuclei}

The origin of light elements during BBN and the existence of the CMB are two successful predictions of the cosmological model, which require the Universe to be hot enough for nuclear reactions to occur in its first few minutes. In this, BBN has been an exceptionally successful theory, describing the formation of light nuclei up to ${}^7$Be in the hot early Universe. It can be considered today as a parameter free theory, insofar the baryon content at present time $\Omega_b h^2 = 0.0224 \pm 0.0001$ and the effective number of relativistic species $N_{\rm eff} = 2.99 \pm 0.17$ --the most relevant cosmological quantities in BBN-- can in principle be determined through the CMB observations
%alone
in combination with baryon acoustic oscillation measurements~\cite{Planck:2018vyg}, and used to make predictions about the primordial light nuclei yields in the specific cosmology considered.

Phenomenologically, BBN can be considered to begin when the weak reactions
\begin{align}
    \label{npeq1}
    p + e^- &\rightleftharpoons  n + \nu_e\\
    \label{npeq2}
    n + e^+ &\rightleftharpoons p + \bar{\nu}_e
\end{align}
fall out of equilibrium at $T \sim 0.7$\,MeV. The neutron-to-proton ratio at the onset of nuclear reactions $n/p$ freezes out except for the decaying free neutrons. At this point the temperature is low enough to allow for the out--of--equilibrium formation of light elements such as ${}^4$He, D, ${}^3$He, up to ${}^7$Be and ${}^7$Li, the latter synthesized at the very cold end of BBN, through a buildup of nuclei incapable of assembling heavier elements~\cite{Iocco:2007km} before nuclear reactions finally freeze. Predictions of the primordial abundances of the light elements are in overall good agreement with those inferred from observational data~\cite{Schramm:1997vs, Steigman:2007xt, Iocco:2008va, Cyburt:2015mya}. For this, BBN has been invoked to firmly constrain deviations from the standard cosmological model below the MeV scale~\cite{Pospelov:2010hj}.

In particular, after the chain of nuclear reactions has started a large fraction of the free neutrons available end up forming helium-4 (${}^4$He). This is expected since ${}^4$He is the nucleus with the highest binding energy between the lightest elements formed during BBN. The primordial mass fraction of ${}^4$He is then expected to amount to around
\begin{equation}
    Y_p = \frac{2(n/p)}{1+n/p} \simeq 0.25\,.
\end{equation}
On the observational side, the helium-4 mass fraction $Y_p$ is inferred through the emission lines of H and He atoms in low-metallicity extragalactic HII regions. According to the most recent review by the Particle Data Group (PDG), the value for the ${}^4$He mass fraction deduced from these measurements is $Y_p = 0.245 \pm 0.003$~\cite{ParticleDataGroup:2022pth}. A smaller fraction of free neutrons eventually ends up into other elements and isotopes, while the fractions of D and ${}^3$He relative to H stand at the level of $\mathcal{O}(10^{-5})$, and ${}^7$Li at the level of about $\mathcal{O}(10^{-10})$. Theoretical arguments for these fractions are found in Refs.~\cite{Esmailzadeh:1990hf, Mukhanov:2003xs}. The fractional abundance of deuterium is measured in gas clouds seen in absorption against an unrelated background source such as a quasar~\cite{1976A&A....50..461A}. The weighted mean of the 11 most precise measurements is $X_D = (2.547 \pm 0.025)\times 10^{-5}$~\cite{ParticleDataGroup:2022pth}

The experimental values are reported in Table~\ref{tab:He4} in comparison with the results from \texttt{PArthENoPE} v3.0~\cite{2022CoPhC.27108205G}. For the choice of main cosmological parameters mentioned in Sec. \ref{sec:results} the code produces results consistent with observations in the standard scenario without PBHs.
\begin{table}[H]
    \centering
    \begin{tabular}{|c|c|c|}
        \hline
        & $X_D \times 10^5$  & $Y_p$ \\
        \hline
        Observations & $2.547 \pm 0.025 $ & $0.245 \pm 0.003$ \\
        \texttt{PArthENoPE} & $2.511 \pm 0.1$ & $0.246 \pm 0.002$ \\
        \hline
    \end{tabular}
    \caption{The weighted value for the observational abundances of deuterium and $^4$He from the PDG~\cite{ParticleDataGroup:2022pth}, compared with the values computed from \texttt{PArthENoPE} v3.0~\cite{2022CoPhC.27108205G}.}
    \label{tab:He4}
\end{table}

In an expanding Universe, the freeze-out of chemical reactions occurs when the reaction rate drops below the expansion rate and equilibrium can no longer be achieved. For this, even a slight modification of the Hubble rate inevitably affects the equilibrium of the ongoing reactions. In the presence of a sizable PBH population, the expansion rate of the Universe can be increased, leading to an earlier freeze-out of the chemical species. This results in an overproduction of neutrons which are the most energetically disfavored species in the reactions in Eqs.~\eqref{npeq1} and~\eqref{npeq2}. Moreover, as the temperature rapidly drops, less neutrons will have decayed at the onset of nuclear reactions. Consequently, the $n/p$ ratio and the mass fraction of ${}^4$He are expected to increase if PBHs in the mass range considered evaporates at the onset of BBN.

The quantity most affecting the reactions of BBN is the baryon-to-photon ratio $\eta = n_b/n_\gamma$, where $n_b$ is the number of baryons and $n_\gamma$ is the number density of photons, per comoving volume.
This ratio is constant in the standard cosmology in absence of interactions that change the number of particles, which is the case at the temperatures considered, at around and below weak freeze-out.
The baryon-to-photon ratio is assessed at recombination to the value determined by the CMB analysis reported by the {\it Planck} collaboration~\cite{Planck:2018vyg}
\begin{equation}
    \label{eq:eta_measurement}
    \eta_{b} = 6.138\times 10^{-10}\,,
\end{equation}
which is the default value used in the analysis by the code \texttt{Parthenope} since there is no change in this quantity between BBN and recombination in standard cosmology, nor in the case considered, since the PBHs evaporate before BBN, and the material injected becomes part of the thermalized cosmological plasma as described. 

In the standard scenario, the value of the baryon-to-photon ratio changes between the beginning and the end of BBN due to the transfer of entropy to photons from $e^+e^-$ annihilation at $T\sim 0.5$\,MeV. The value of $\eta$ before the onset of BBN, $\eta_{i}$, is then obtained from its value at recombination as in Ref.~\cite{Serpico:2004gx}
\begin{equation}
    \label{eq:eta_standard}
    \eta_{i} = \frac{g_s(T_{i})}{g_s(T_{\rm rec})} \, \eta_{b} \simeq 2.75\, \eta_b \,,
\end{equation}
where $T_{\rm rec} \sim\,$eV and $T_{i}$ is conventionally set in the code \texttt{PArthENoPE} to 10\,MeV.

In the scenario we explore, PBHs evaporating around the time of $e^+e^-$ annihilation constitute an additional source of entropy in the plasma, and it is indeed crucial to estimate their effect on the alteration of the baryon-to-photon ratio. To do so it is useful to express $\eta$ in terms of the entropy density,
\begin{equation}
\label{eq:etabs}
    \eta \propto g_s(T)\frac{n_b}{s} \,,
\end{equation}
we neglect the baryon contribution from PBHs evaporation, so that the evolution of $n_b$ is given by
\begin{equation}
\label{eq:nbar}
    \dot{n_b} + 3Hn_b = 0\,.
\end{equation}
Combining Eqs.~\eqref{eq:etabs} and ~\eqref{eq:nbar} with Eq.~\eqref{eq:sT} we find
\begin{equation}
    \label{eq:etabaryon}
    \dot\eta = \frac{3}{4}\frac{\dot{M}_{\rm PBH}}{M_{\rm PBH}} \frac{\rho_{\rm PBH}}{\rho_r} \,\eta\,,
\end{equation}
which can be numerically solved together with Eq.~\eqref{eq:PBHabundance} to obtain the evolution of $\eta$ for different values of the parameters $(\beta, M_{\rm PBH})$.
The deuterium abundance is directly related to the baryon-to-photon ratio $\eta$ so that a change in $\eta$ greatly affects the fraction $X_D$. Deuterium formation begins at temperatures much lower than its binding energy $\Delta_D \simeq 2.2$\,MeV due to dissociation by high energy photons. The condition for deuterium to form is that the number of photons with an energy above $\Delta_D$, per baryon, which is roughly $\eta^{-1} e^{-\Delta_D/T}$, falls below unity. In the standard scenario this happens at $T \sim 0.1$\,MeV. In the presence of a higher baryon-to-photon ratio, deuterium production is anticipated compared to the standard case and so is the production of all the other nuclei, being the deuterium necessary to start the chain of reactions. An earlier production of deuterium leads to its more efficient conversion into heavier elements, ultimately resulting in a lower primordial abundance with respect to the standard case. Conversely, a lower baryon-to-photon ratio induces a later onset of deuterium synthesis, a less efficient build-up of heavier elements and consequently a higher deuterium abundance at the end of BBN.

Given the same initial conditions, and {\it ab initio} baryon-to-photon ratio, a Universe filled with PBH will experience a reduced baryon-to-photon ratio (with respect to the standard case) after their evaporation.
 This is because the higher entropy per baryon of the particles injected lowers $\eta$.
 In this sense, it can be said that PBH evaporation would increase the entropy of the Universe and hence induce a higher deuterium abundance. 
 If one adopts the opposite point of view and imposes instead that the value of $\eta$ at the end of BBN is the same as what is measured through the CMB --which must be the case if no exotic physics takes place in the Universe after PBH evaporation-- then the baryon-to-photon ratio of the Universe pre--existing the PBH evaporation must be higher than in the standard case (the intervening PBH evaporation lowers it to the CMB value). This in turn forces an earlier start of the deuterium nucleosynthesis, which in turn leads to a lower final abundance. This is the approach we adopt here, and in Figs.~\ref{fig:Yp} and~\ref{fig:D} we show the BBN abundance of deuterium and helium-4 as a function of increasing PBH abundance at formation. The effect on deuterium described above is what drives the constraints we derive\footnote{We have also tested the effect of modification of the Hubble rate, described at the beginning of this section, by artificially (and unphysically) switching on only one of the two at the time: as intuition would have it we do see that the Hubble rate modification alone has a leading effect on helium, whereas the baryon-to-photon ratio affects both helium and deuterium, with the latter undergoing the most dramatic modification and thus leading the resulting constraints.}.

\section{Results}
\label{sec:results}
BBN is a strongly non-linear process in which multiple reactions take place at the same time. For this, numerical methods for the integration of coupled differential equations are required in order to obtain an estimate of its final outcome. For such calculations we employ here the numerical code \texttt{PArthENoPE} v3.0~\cite{2022CoPhC.27108205G}, which allows us to consider up to 26 nuclides and 100 reactions to solve for the whole system of differential equations and compute reliable theoretical predictions for the light nuclei abundances. While the set of BBN reactions solved by the code \texttt{PArthENoPE} is not modified, we account for the altered expansion history of the Universe by solving for Eqs.~\eqref{eq:PBHabundance} and~\eqref{eq:etabaryon} numerically for given values of the parameters $\beta$ and $M_{\rm PBH}$ and with the initial condition in Eq.~\eqref{beta}, following the steps detailed in previous work~\cite{Masina:2020xhk, Mazde:2022sdx}. The resulting Hubble rate $H(T)$ as given in Eq.~\eqref{hub} is then treated as an input for the background evolution over which the coupled BBN equations are solved with the FORTRAN77 source code.

The value of the baryon-to-photon ratio at the onset of BBN is related to its value at recombination through Eq.~\eqref{eq:eta_standard} in \texttt{PArthENoPE}. We account for the modified evolution of $\eta$ due to the entropy injection by PBHs by changing the default value in the source code to the value of $\eta_{i}$ obtained from the solution of Eq.~\eqref{eq:etabaryon} with the condition $\eta = \eta_b$ at recombination. The distribution of the PBH masses is assumed to be monochromatic and centered around the value $M_{\rm PBH}$.

\begin{figure}[htbp!]
    \centering
    \includegraphics[width = \linewidth]{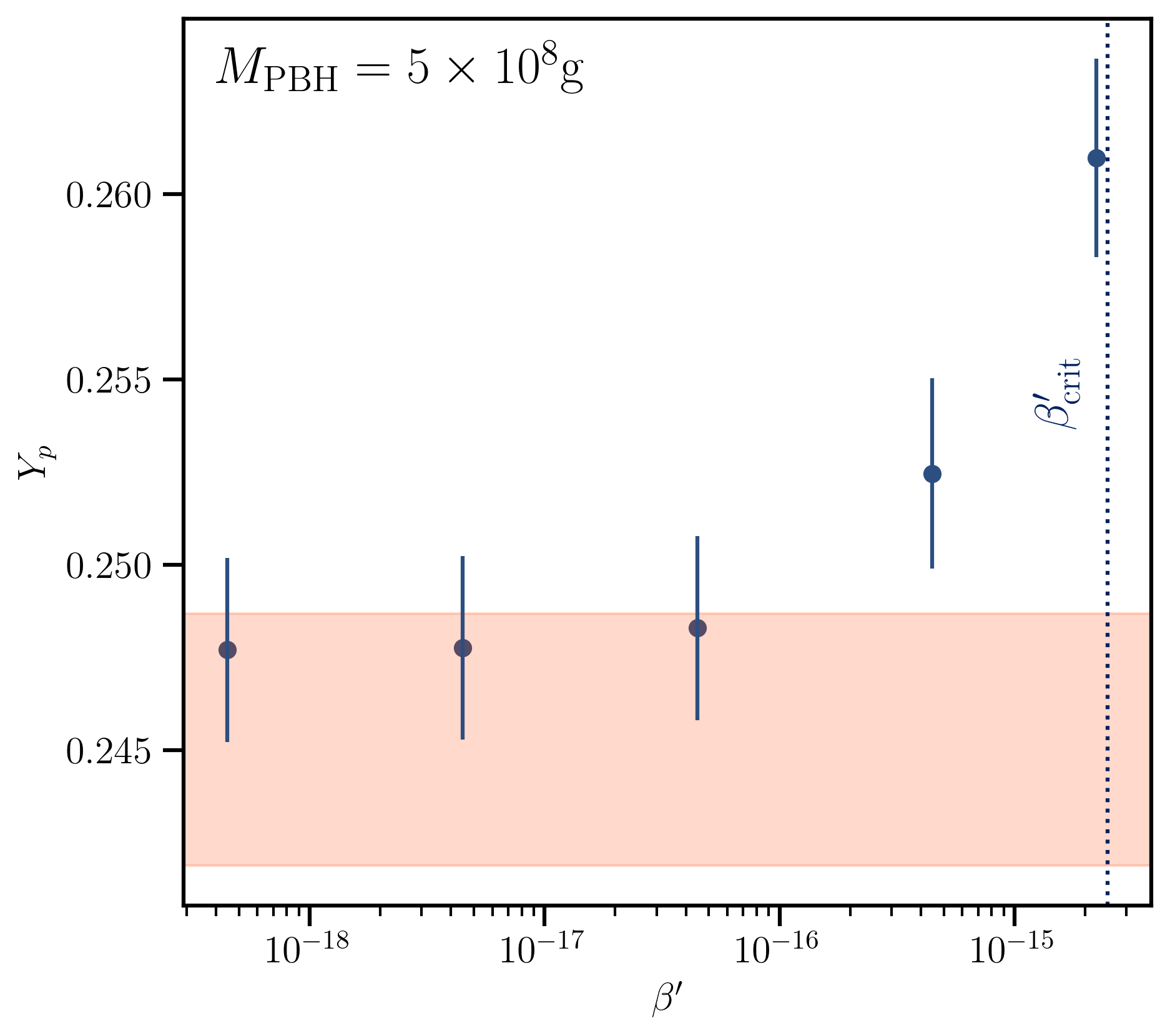}
    \caption{Primordial mass fraction of ${}^4$He for different values of the parameter $\beta'$, for a population of PBHs with M$_{\rm PBH} = 5 \times 10^8$\,g and a monochromatic mass distribution. Also shown is the measurement for $Y_p$ reported by the PDG~\cite{ParticleDataGroup:2022pth}, see the red band giving the 1$\sigma$ range.}
    \label{fig:Yp}
\end{figure}

\begin{figure}[htbp!]
    \centering
    \includegraphics[width=0.45\textwidth]{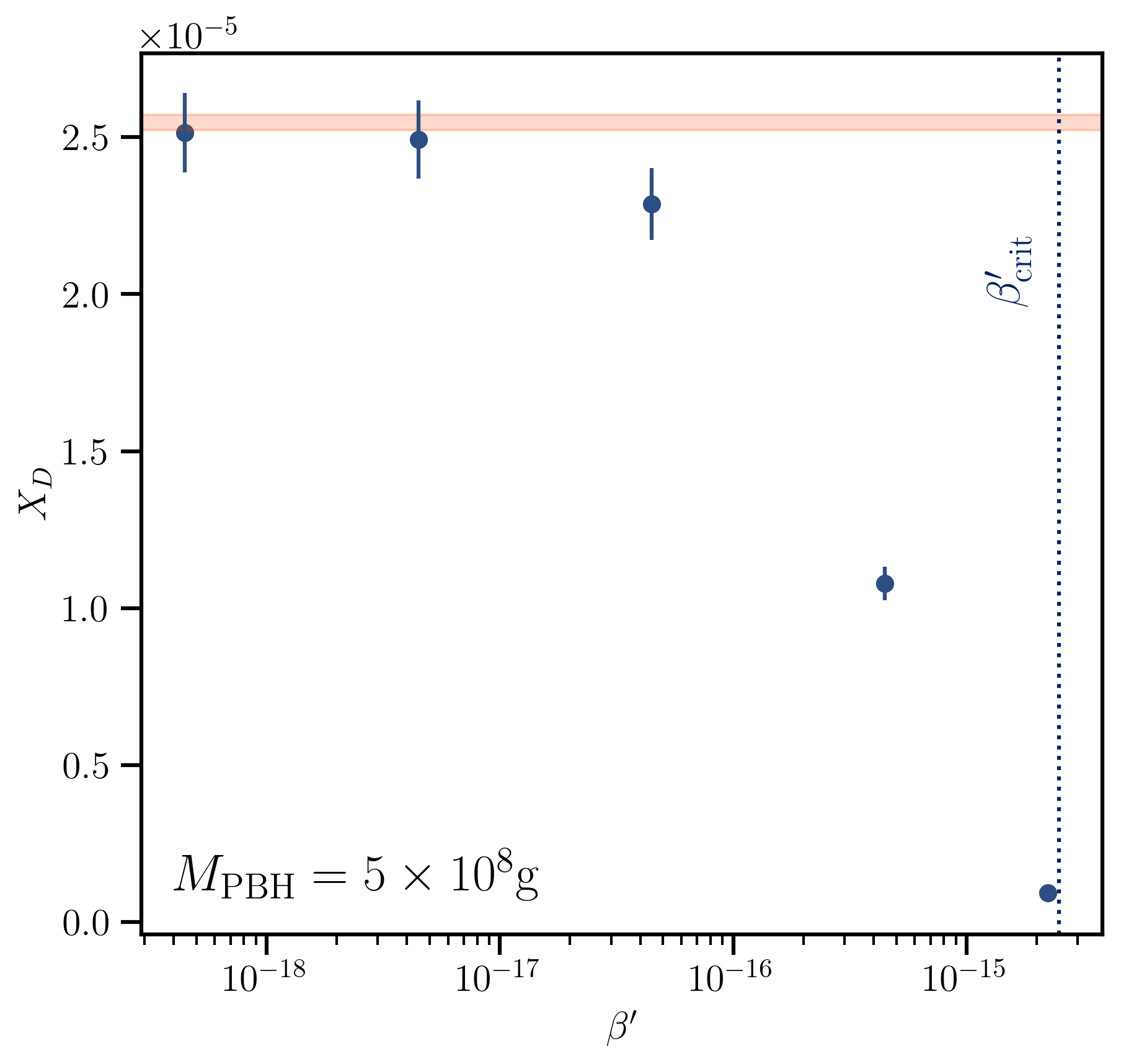}
    \caption{Primordial abundance of deuterium for different values of the parameter $\beta'$, for a population of PBHs with M$_{\rm PBH} = 5 \times 10^8$\,g. Also shown is the measurement for $X_D$ reported by the PDG~\cite{ParticleDataGroup:2022pth}, see the red band giving the 1$\sigma$ range.}
    \label{fig:D}
\end{figure}

Figure~\ref{fig:Yp} shows the primordial mass fraction of ${}^4$He --$Y_p$ as customary in literature-- as a function of the parameter $\beta'$ in Eq.~\eqref{eq:betaprime}. The result is obtained assuming a population of PBHs with a monochromatic mass distribution centered around M$_{\rm PBH} = 5 \times 10^8$\,g. As the relative population of PBHs increases with increasing values of $\beta'$, the mass fraction $Y_p$ starts growing significantly due to the modified cosmology impacting around the onset of BBN. We have explored the region of parameters around the critical value $\beta'_{\rm crit}$ related to the quantity in Eq.~\eqref{eq:betacrit} and marked by the vertical dotted line. Once the value of $Y_p$ for given $\beta'$ is determined through \texttt{PArthENoPE}, the error bar is associated assuming a fractional error of 1\%~\cite{ParticleDataGroup:2022pth}. In Fig.~\ref{fig:Yp}, the red band gives the 1$\sigma$ range associated with the mass fraction from PDG, see Table~\ref{tab:He4}. Similarly, the fraction of deuterium $X_D$ is shown in Fig.~\ref{fig:D} as a function of the PBH primordial abundance $\beta'$. As explained below Eq.~\eqref{eq:etabaryon}, an increase in the PBH abundance leads to a decrease in $X_D$ due to the anticipated production of deuterium for any given PBH mass.

As described in the previous section, the deuterium fraction is more sensible to the increasing $\beta'$ with respect to helium, so it provides more stringent bounds.

Using the methods described above, we compare the results for the primordial abundance of light elements, obtained for different values of $(\beta, M_{\rm PBH})$, with the observational constraints on the abundances of the elements ${}^4$He and D. As we discussed, helium and deuterium fractions are extremely sensible to the physical alterations introduced at BBN. The code \texttt{PArthENoPE} v3.0 is run across the range of values $10^8{\rm\,g} \leq M_{\rm PBH} \leq 10^9{\rm\,g}$ and $10^{-18} \leq \beta \leq 10^{-14}$. For a given value of the PBH mass population, the bound on $\beta'$ is obtained by comparing the value for $Y_p$ and $X_D$ obtained experimentally, red band in Fig.~\ref{fig:Yp} and Fig.~\ref{fig:D}, with those determined through \texttt{PArthENoPE} v3.0. We request that our predictions from the model does not deviate from what is reported by PDG by more than 2 sigma or 95\% confidence level or CL, where the uncertainties for the values of $Y_p$ and $X_D$ are inferred from \texttt{PArthENoPE}, and reported by observations are added in squares. 

 We do not include other light elements produced during BBN in our analysis, as their primordial abundances are suppressed with respect to deuterium and helium, and they are further affected by the subsequent stellar evolution, which contributes to additional uncertainties in the determination of their primordial abundances and are thus not to be used for cosmology purposes \cite{Iocco:2008va, Iocco:2012vg}. The default value for the baryon-to-photon ratio at recombination used in the analysis is given in Eq.~\eqref{eq:eta_measurement}, while the neutron lifetime is set at the value $\tau_n = (879.4\pm0.6)$\,s~\cite{10.1093/ptep/ptaa104}.

\begin{figure}[H]
    \centering
    \includegraphics[width = \linewidth]{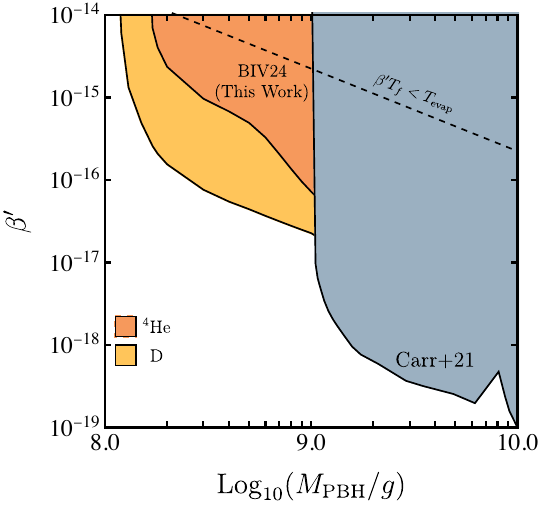}
    \caption{The bounds at 95\% confidence level on PBHs in the mass range $10^8{\rm\,g} \leq M_{\rm PBH} \leq 10^9{\rm\,g}$ for the helium abundance (orange shaded area) and deuterium (okra shaded area). Also shown are the results from Ref.~\cite{Carr:2020gox} (gray shaded area).}
    \label{fig:bounds}
\end{figure}

Figure~\ref{fig:bounds} returns the upper bounds on $\beta'(M)$ at 95\% CL obtained from our methods of considering the effects of a modified expansion rate by the presence of PBHs at the onset of BBN. This is shown by the area in orange shade labeled ``BIV24 (This work)''. We have considered the effects resulting from the modifications in the helium abundance alone (orange shaded area) and deuterium (okra shaded area). PBHs of mass $M_{\rm PBH} \lesssim 10^8$\,g evaporate too early to affect the evolution of BBN.

On the opposite end, PBHs heavier than about $10^9$\,g would evaporate well into the BBN epoch, so that the by-products of their evaporation could affect the nuclear reactions. While we do not consider these effects in our analysis, much tighter constraints can be drawn from including these by-products as discussed in Ref.~\cite{Carr:2020gox}, see the gray shaded area labeled ``Carr+21''.

\section{Discussion}
\label{sec:discussion}

By accounting for the physics described above, we can constrain the abundance of PBHs at their formation $\beta$ in the mass range $10^8{\rm\,g} \lesssim M_{\rm PBH} \lesssim 10^9{\rm\,g}$. 
In the mass region of interest, our results improve the bounds on $\beta$ by several orders of magnitude with respect to previous analyses present in the literature, which account for the secondary gravitational wave produced after PBH evaporation and their effects on the cosmic microwave background~\cite{Domenech:2021wkk}.
Given that for PBH masses considered here the evaporation process ends well before the formation of nuclei, $T_{\rm evap} \gtrsim 1 \, \rm MeV $, we have deliberately omitted the non-thermal effects of high energy particles injection on nucleosynthesis. We have estimated that all high energy particles emitted by the evaporation thermalize ``on--the--spot'', namely within a timescale much shorter than the Hubble one at temperature $T$. For proton-proton SM interactions, thermalization can be regarded as instantaneous at the moment (and therefore cosmic temperature) of evaporation, which for the masses considered is before the very onset of BBN, as recalled above. This avoids any dissociation of nuclei by highly energetic photons and hadrons, and allows us to remain within the well-based approximation of thermal nucleosynthesis.

In principle, our results could be affected by the emission of pions, kaons, and anti-nucleons, which lose their initial kinetic energy through elastic interactions, such as Coulomb and Compton scattering, thereby preventing the thermal equilibrium to be reached. However, mesons scatter much more efficiently off the abundant photons of the pre-existing background than off neutrons or protons, leading to their rapid decay and thermalization, well before any significant effects on light elements occur~\cite{Kohri:1999ex}. As a result, the neutrons and protons remain thermalized and form part of the background, with their neutron-to-proton ratio determined solely by the temperature, as in standard BBN.

%In the analysis, we have neglected the effects brought in by the flux of neutrinos emitted by PBHs before or during neutrino decoupling, which leads to a modification of the abundance of deuterium~\cite{1978SvA....22..138V}. We also expect a change in the effective number of relativistic neutrinos $N_{\rm eff}$ and indirectly on $Y_p$, leading to another complementary effect in addition to our results~\cite{Mastrototaro:2021wzl}. In such a scenario, our results should be regarded as conservative. These aspects will be treated in future work.

In the analysis, we have neglected the effects brought in by the flux of neutrinos emitted by PBHs before or during neutrino decoupling, which leads to a modification of the abundance of deuterium~\cite{1978SvA....22..138V}. We also expect a change in the effective number of relativistic neutrinos $N_{\rm eff}$ and indirectly on $Y_p$, leading to another effect in addition to our results~\cite{Mastrototaro:2021wzl}. In this scenario, a lower value for the effective number of relativistic neutrinos $N_{\rm eff} < 3$ would partially compensate the increase in the Hubble rate. Both deuterium and helium abundances would be affected, with the most relevant effect on helium, as deuterium is less sensitive to changes in the Hubble rate. While this effect certainly deserves a thorough study, it remains subleading in the current case. The most stringent constraints arise from deuterium abundance, which is highly sensitive to the entropy injection provided by PBH evaporation, whereas the effects from the Hubble rate are subleading.

%Despite the effects on Helium and Deuterium, the change in the Hubble rate induced by the presence of a PBH population, together with entropy immision, tightens the bounds as in Fig. \ref{fig:bounds}, leading to an effect that has to be taken in account in future publications.}

In Fig.~\ref{fig:bounds} we have compared our results with the previous work in Refs.~\cite{Carr:2009jm, Carr:2020gox}, quoting the published numbers at face-value. In doing this, we have relied on the expression for the evaporation rate of PBHs in Eq.~\eqref{eq:Mdot}, from which the time of the evaporation in Eq.~\eqref{eq:tevap} is derived. In our estimates from Eq.~\eqref{eq:tevap}, the PBH mass for the reheating bound (i.e the lowest that preserves the Universe well-behaving into radiation domination by the reheating temperature $T_{\rm RH} \approx 4$\,MeV  ~\cite{Kawasaki:1999na, Kawasaki:2000en,deSalas:2015glj, Hasegawa:2019jsa}) is $M_{\rm PBH} \approx 5\times10^8\,$g.
This mass value differs from what is used in Refs.~\cite{Carr:2009jm, Carr:2020gox} which, for the reheating temperature $T_{\rm RH} \approx 4$\,MeV, obtain the PBH mass $M_{\rm PBH} = 10^9\,$g. Using the formula for $t_{\rm evap}$ in Eq.~\eqref{eq:tevap}, the gray region constrained in Fig.~\ref{fig:bounds} would shift to lower values of the mass and partially overlap with our results. Two considerations are in order here in this regard:
{\it a)} the reheating bound only applies when PBH density is such that their evaporation is able to reheat the Universe, a condition that we are staying far away from (PBH densities considered here are way lower);
{\it b)} the approximation we adopt is more complete, and therefore we believe that all current bounds should be reassessed by including the effects from particle emission studied elsewhere~\cite{Kawasaki:2004qu, Kawasaki:2017bqm, Kawasaki:2020qxm}. Nevertheless, this likely only implies a minor quantitative shift along the mass axes, and no qualitatively alteration of the results. Studying its consequences is beyond the scope of this paper, and we leave it for further study.

Furthermore, our results are derived under the assumption of a monochromatic mass spectrum for the PBH distribution. However, the mass distribution could span various decades of mass as it is expected in models where PBH formation possesses a low mass tail from critical collapse~\cite{Niemeyer:1997mt}. Accounting for an extended mass distribution is generally non trivial and requires both an improvement of the methods presented in Sec.~\ref{sec:methods} and further investigations on the distribution to be used. Additional phenomenology, such as accretion and merging of PBHs, may arise leading to the necessity of a larger set of parameters. In any case with a non-monochromatic mass distribution we expect the impact on the Hubble rate to be weaker but to last longer due to the higher masses tail of the distribution. This may affect our constraints loosening them, simultaneously broadening them across a wider mass spectrum, particularly towards lower masses. The last statement applies in the case of a narrow mass distribution, for which the definition of $\beta$ can be obtained by integrating the explicit mass distribution over the whole mass range. For a wider mass distribution, the time dependence of the width of the distribution should be taken into account due to the different evaporation times between the lower and higher masses. The approach usually adopted for this case is the binning of the distribution into intervals of size $\Delta M_{\rm PBH} \sim M_{\rm PBH}$ which are treated separately.

\section{Conclusions}
\label{sec:conclusions}

We examined the effects of primordial black holes as they undergo evaporation during the weak freeze-out phase at the onset of Primordial Nucleosynthesis (BBN). By solving the differential equations governing PBHs and radiation density evolution alongside the Friedmann equations, we were able to model the evolution of the background cosmology in the presence of PBHs of various masses and abundance at formation, assuming a monochromatic mass spectrum. We have accounted for the change in the baryon-to-photon ratio and Hubble rate that are brought forth by the PBH evaporation.
We have implemented these results into the BBN numerical code \texttt{PArthENoPE}, enabling us to compute the primordial abundance of light nuclei as a function of the parameters $(\beta, M_{\rm PBH})$.

Our analysis shows significant deviations in the abundance of deuterium and ${}^4$He from the standard, no--PBH case, for values of $\beta$ approaching the critical threshold $\beta_{\rm crit} \equiv T_{\rm evap}/T_f$, for PBHs masses within the range $10^8{\rm\,g} \leq M_{\rm PBH} \leq 10^9{\rm\,g}$. 
By comparing the abundances of deuterium and ${}^4$He theoretically predicted in each PBH scenario against the empirically observed primordial abundances, we have derived the constraints in the PBH parameters.
Our results request the parameter $\beta'$ to be lower than the critical value $\beta_{\rm crit}$ for the whole range of masses considered, thereby ruling out the possibility of early PBH domination. The strongest bound are obtained for the heaviest PBHs considered, with $\beta' \lesssim 10^{-16}$ for $M_{\rm PBH} = 10^9$\,g.

This analysis yields novel constraints within the parameter space of PBHs, more stringent with respect to previous results over the same mass range. The bounds obtained with the method presented here could be strengthened by considering the effects of high energy neutrinos emission before neutrino decoupling that could impact $N_{\rm eff}$. Future work should be devoted to a more complete treatment of these effects of evaporation products and should extend the analysis to other mass distributions and models.

\vspace{1cm}

\begin{acknowledgments}
We thank O.\ Pisanti for help with the \texttt{PArthENoPE} code, P.\ D.\ Serpico for useful comments on the manuscript, M.\ Zantedeschi for the insight provided on the phenomenology of PBHs evaporation, and K.\ Kohri for stimulating conversations. 

F.I.\ is partially supported by the research grant No.\ 2022E2J4RK ``PANTHEON: Perspectives in Astroparticle and Neutrino THEory with Old and New messengers'' under the program PRIN 2022 funded by the Italian Ministero dell’Universit\`a e della Ricerca (MUR) and by the European Union – Next Generation EU.
A.B.\ and F.I.\ acknowledge the hospitality of the Tsung-Dao Lee Institute (TDLI) in Shanghai (China) and Universit\`a di Ferrara (Italy) during the final phases of this work, funded by the TAsP Iniziativa Specifica of INFN.
L.V.\ acknowledges support by the National Natural Science Foundation of China (NSFC) through the grant No.\ 12350610240 ``Astrophysical Axion Laboratories'', as well as hospitality by the Istituto Nazionale di Fisica Nucleare (INFN) Frascati National Laboratories, the Galileo Galilei Institute for Theoretical Physics in Firenze (Italy), the INFN section of Napoli (Italy), the INFN section of Trento (Italy), the INFN section of Ferrara (Italy), and the University of Texas at Austin (USA) throughout the completion of this work. This publication is based upon work from the COST Actions ``COSMIC WISPers'' (CA21106) and ``Addressing observational tensions in cosmology with systematics and fundamental physics (CosmoVerse)'' (CA21136), both supported by COST (European Cooperation in Science and Technology).
\end{acknowledgments}

\bibliography{Bibliography.bib}

\begin{thebibliography}{50}%
\makeatletter
\providecommand \@ifxundefined [1]{%
 \@ifx{#1\undefined}
}%
\providecommand \@ifnum [1]{%
 \ifnum #1\expandafter \@firstoftwo
 \else \expandafter \@secondoftwo
 \fi
}%
\providecommand \@ifx [1]{%
 \ifx #1\expandafter \@firstoftwo
 \else \expandafter \@secondoftwo
 \fi
}%
\providecommand \natexlab [1]{#1}%
\providecommand \enquote  [1]{``#1''}%
\providecommand \bibnamefont  [1]{#1}%
\providecommand \bibfnamefont [1]{#1}%
\providecommand \citenamefont [1]{#1}%
\providecommand \href@noop [0]{\@secondoftwo}%
\providecommand \href [0]{\begingroup \@sanitize@url \@href}%
\providecommand \@href[1]{\@@startlink{#1}\@@href}%
\providecommand \@@href[1]{\endgroup#1\@@endlink}%
\providecommand \@sanitize@url [0]{\catcode `\\12\catcode `\$12\catcode `\&12\catcode `\#12\catcode `\^12\catcode `\_12\catcode `\%12\relax}%
\providecommand \@@startlink[1]{}%
\providecommand \@@endlink[0]{}%
\providecommand \url  [0]{\begingroup\@sanitize@url \@url }%
\providecommand \@url [1]{\endgroup\@href {#1}{\urlprefix }}%
\providecommand \urlprefix  [0]{URL }%
\providecommand \Eprint [0]{\href }%
\providecommand \doibase [0]{http://dx.doi.org/}%
\providecommand \selectlanguage [0]{\@gobble}%
\providecommand \bibinfo  [0]{\@secondoftwo}%
\providecommand \bibfield  [0]{\@secondoftwo}%
\providecommand \translation [1]{[#1]}%
\providecommand \BibitemOpen [0]{}%
\providecommand \bibitemStop [0]{}%
\providecommand \bibitemNoStop [0]{.\EOS\space}%
\providecommand \EOS [0]{\spacefactor3000\relax}%
\providecommand \BibitemShut  [1]{\csname bibitem#1\endcsname}%
\let\auto@bib@innerbib\@empty
%</preamble>
\bibitem [{\citenamefont {Zel'dovich}\ and\ \citenamefont {Novikov}(1967)}]{Zeldovich:1967lct}%
  \BibitemOpen
  \bibfield  {author} {\bibinfo {author} {\bibfnamefont {Y.~B.}\ \bibnamefont {Zel'dovich}}\ and\ \bibinfo {author} {\bibfnamefont {I.~D.}\ \bibnamefont {Novikov}},\ }\href@noop {} {\bibfield  {journal} {\bibinfo  {journal} {Sov. Astron.}\ }\textbf {\bibinfo {volume} {10}},\ \bibinfo {pages} {602} (\bibinfo {year} {1967})}\BibitemShut {NoStop}%
\bibitem [{\citenamefont {Hawking}(1971)}]{Hawking:1971ei}%
  \BibitemOpen
  \bibfield  {author} {\bibinfo {author} {\bibfnamefont {S.}~\bibnamefont {Hawking}},\ }\href {\doibase 10.1093/mnras/152.1.75} {\bibfield  {journal} {\bibinfo  {journal} {Mon. Not. Roy. Astron. Soc.}\ }\textbf {\bibinfo {volume} {152}},\ \bibinfo {pages} {75} (\bibinfo {year} {1971})}\BibitemShut {NoStop}%
\bibitem [{\citenamefont {Carr}\ and\ \citenamefont {Hawking}(1974)}]{Carr:1974nx}%
  \BibitemOpen
  \bibfield  {author} {\bibinfo {author} {\bibfnamefont {B.~J.}\ \bibnamefont {Carr}}\ and\ \bibinfo {author} {\bibfnamefont {S.~W.}\ \bibnamefont {Hawking}},\ }\href {\doibase 10.1093/mnras/168.2.399} {\bibfield  {journal} {\bibinfo  {journal} {Mon. Not. Roy. Astron. Soc.}\ }\textbf {\bibinfo {volume} {168}},\ \bibinfo {pages} {399} (\bibinfo {year} {1974})}\BibitemShut {NoStop}%
\bibitem [{\citenamefont {Carr}(1975)}]{Carr:1975qj}%
  \BibitemOpen
  \bibfield  {author} {\bibinfo {author} {\bibfnamefont {B.~J.}\ \bibnamefont {Carr}},\ }\href {\doibase 10.1086/153853} {\bibfield  {journal} {\bibinfo  {journal} {Astrophys. J.}\ }\textbf {\bibinfo {volume} {201}},\ \bibinfo {pages} {1} (\bibinfo {year} {1975})}\BibitemShut {NoStop}%
\bibitem [{\citenamefont {Shibata}\ and\ \citenamefont {Sasaki}(1999)}]{Shibata:1999zs}%
  \BibitemOpen
  \bibfield  {author} {\bibinfo {author} {\bibfnamefont {M.}~\bibnamefont {Shibata}}\ and\ \bibinfo {author} {\bibfnamefont {M.}~\bibnamefont {Sasaki}},\ }\href {\doibase 10.1103/PhysRevD.60.084002} {\bibfield  {journal} {\bibinfo  {journal} {Phys. Rev. D}\ }\textbf {\bibinfo {volume} {60}},\ \bibinfo {pages} {084002} (\bibinfo {year} {1999})},\ \Eprint {http://arxiv.org/abs/gr-qc/9905064} {arXiv:gr-qc/9905064} \BibitemShut {NoStop}%
\bibitem [{\citenamefont {Niemeyer}\ and\ \citenamefont {Jedamzik}(1999)}]{Niemeyer:1999ak}%
  \BibitemOpen
  \bibfield  {author} {\bibinfo {author} {\bibfnamefont {J.~C.}\ \bibnamefont {Niemeyer}}\ and\ \bibinfo {author} {\bibfnamefont {K.}~\bibnamefont {Jedamzik}},\ }\href {\doibase 10.1103/PhysRevD.59.124013} {\bibfield  {journal} {\bibinfo  {journal} {Phys. Rev. D}\ }\textbf {\bibinfo {volume} {59}},\ \bibinfo {pages} {124013} (\bibinfo {year} {1999})},\ \Eprint {http://arxiv.org/abs/astro-ph/9901292} {arXiv:astro-ph/9901292} \BibitemShut {NoStop}%
\bibitem [{\citenamefont {Musco}\ \emph {et~al.}(2005)\citenamefont {Musco}, \citenamefont {Miller},\ and\ \citenamefont {Rezzolla}}]{Musco:2004ak}%
  \BibitemOpen
  \bibfield  {author} {\bibinfo {author} {\bibfnamefont {I.}~\bibnamefont {Musco}}, \bibinfo {author} {\bibfnamefont {J.~C.}\ \bibnamefont {Miller}}, \ and\ \bibinfo {author} {\bibfnamefont {L.}~\bibnamefont {Rezzolla}},\ }\href {\doibase 10.1088/0264-9381/22/7/013} {\bibfield  {journal} {\bibinfo  {journal} {Class. Quant. Grav.}\ }\textbf {\bibinfo {volume} {22}},\ \bibinfo {pages} {1405} (\bibinfo {year} {2005})},\ \Eprint {http://arxiv.org/abs/gr-qc/0412063} {arXiv:gr-qc/0412063} \BibitemShut {NoStop}%
\bibitem [{\citenamefont {Carr}\ \emph {et~al.}(2021{\natexlab{a}})\citenamefont {Carr}, \citenamefont {Kuhnel},\ and\ \citenamefont {Visinelli}}]{Carr:2020erq}%
  \BibitemOpen
  \bibfield  {author} {\bibinfo {author} {\bibfnamefont {B.}~\bibnamefont {Carr}}, \bibinfo {author} {\bibfnamefont {F.}~\bibnamefont {Kuhnel}}, \ and\ \bibinfo {author} {\bibfnamefont {L.}~\bibnamefont {Visinelli}},\ }\href {\doibase 10.1093/mnras/staa3651} {\bibfield  {journal} {\bibinfo  {journal} {Mon. Not. Roy. Astron. Soc.}\ }\textbf {\bibinfo {volume} {501}},\ \bibinfo {pages} {2029} (\bibinfo {year} {2021}{\natexlab{a}})},\ \Eprint {http://arxiv.org/abs/2008.08077} {arXiv:2008.08077 [astro-ph.CO]} \BibitemShut {NoStop}%
\bibitem [{\citenamefont {Carr}\ \emph {et~al.}(2016)\citenamefont {Carr}, \citenamefont {Kuhnel},\ and\ \citenamefont {Sandstad}}]{Carr:2016drx}%
  \BibitemOpen
  \bibfield  {author} {\bibinfo {author} {\bibfnamefont {B.}~\bibnamefont {Carr}}, \bibinfo {author} {\bibfnamefont {F.}~\bibnamefont {Kuhnel}}, \ and\ \bibinfo {author} {\bibfnamefont {M.}~\bibnamefont {Sandstad}},\ }\href {\doibase 10.1103/PhysRevD.94.083504} {\bibfield  {journal} {\bibinfo  {journal} {Phys. Rev. D}\ }\textbf {\bibinfo {volume} {94}},\ \bibinfo {pages} {083504} (\bibinfo {year} {2016})},\ \Eprint {http://arxiv.org/abs/1607.06077} {arXiv:1607.06077 [astro-ph.CO]} \BibitemShut {NoStop}%
\bibitem [{\citenamefont {Green}\ and\ \citenamefont {Kavanagh}(2021)}]{Green:2020jor}%
  \BibitemOpen
  \bibfield  {author} {\bibinfo {author} {\bibfnamefont {A.~M.}\ \bibnamefont {Green}}\ and\ \bibinfo {author} {\bibfnamefont {B.~J.}\ \bibnamefont {Kavanagh}},\ }\href {\doibase 10.1088/1361-6471/abc534} {\bibfield  {journal} {\bibinfo  {journal} {J. Phys. G}\ }\textbf {\bibinfo {volume} {48}},\ \bibinfo {pages} {043001} (\bibinfo {year} {2021})},\ \Eprint {http://arxiv.org/abs/2007.10722} {arXiv:2007.10722 [astro-ph.CO]} \BibitemShut {NoStop}%
\bibitem [{\citenamefont {Duechting}(2004)}]{Duechting:2004dk}%
  \BibitemOpen
  \bibfield  {author} {\bibinfo {author} {\bibfnamefont {N.}~\bibnamefont {Duechting}},\ }\href {\doibase 10.1103/PhysRevD.70.064015} {\bibfield  {journal} {\bibinfo  {journal} {Phys. Rev. D}\ }\textbf {\bibinfo {volume} {70}},\ \bibinfo {pages} {064015} (\bibinfo {year} {2004})},\ \Eprint {http://arxiv.org/abs/astro-ph/0406260} {arXiv:astro-ph/0406260} \BibitemShut {NoStop}%
\bibitem [{\citenamefont {Hawking}(1975)}]{Hawking:1975vcx}%
  \BibitemOpen
  \bibfield  {author} {\bibinfo {author} {\bibfnamefont {S.~W.}\ \bibnamefont {Hawking}},\ }\href {\doibase 10.1007/BF02345020} {\bibfield  {journal} {\bibinfo  {journal} {Commun. Math. Phys.}\ }\textbf {\bibinfo {volume} {43}},\ \bibinfo {pages} {199} (\bibinfo {year} {1975})},\ \bibinfo {note} {[Erratum: Commun.Math.Phys. 46, 206 (1976)]}\BibitemShut {NoStop}%
\bibitem [{\citenamefont {Kawasaki}\ \emph {et~al.}(2005)\citenamefont {Kawasaki}, \citenamefont {Kohri},\ and\ \citenamefont {Moroi}}]{Kawasaki:2004qu}%
  \BibitemOpen
  \bibfield  {author} {\bibinfo {author} {\bibfnamefont {M.}~\bibnamefont {Kawasaki}}, \bibinfo {author} {\bibfnamefont {K.}~\bibnamefont {Kohri}}, \ and\ \bibinfo {author} {\bibfnamefont {T.}~\bibnamefont {Moroi}},\ }\href {\doibase 10.1103/PhysRevD.71.083502} {\bibfield  {journal} {\bibinfo  {journal} {Phys. Rev. D}\ }\textbf {\bibinfo {volume} {71}},\ \bibinfo {pages} {083502} (\bibinfo {year} {2005})},\ \Eprint {http://arxiv.org/abs/astro-ph/0408426} {arXiv:astro-ph/0408426} \BibitemShut {NoStop}%
\bibitem [{\citenamefont {Kawasaki}\ \emph {et~al.}(2018)\citenamefont {Kawasaki}, \citenamefont {Kohri}, \citenamefont {Moroi},\ and\ \citenamefont {Takaesu}}]{Kawasaki:2017bqm}%
  \BibitemOpen
  \bibfield  {author} {\bibinfo {author} {\bibfnamefont {M.}~\bibnamefont {Kawasaki}}, \bibinfo {author} {\bibfnamefont {K.}~\bibnamefont {Kohri}}, \bibinfo {author} {\bibfnamefont {T.}~\bibnamefont {Moroi}}, \ and\ \bibinfo {author} {\bibfnamefont {Y.}~\bibnamefont {Takaesu}},\ }\href {\doibase 10.1103/PhysRevD.97.023502} {\bibfield  {journal} {\bibinfo  {journal} {Phys. Rev. D}\ }\textbf {\bibinfo {volume} {97}},\ \bibinfo {pages} {023502} (\bibinfo {year} {2018})},\ \Eprint {http://arxiv.org/abs/1709.01211} {arXiv:1709.01211 [hep-ph]} \BibitemShut {NoStop}%
\bibitem [{\citenamefont {Kawasaki}\ \emph {et~al.}(2020)\citenamefont {Kawasaki}, \citenamefont {Kohri}, \citenamefont {Moroi}, \citenamefont {Murai},\ and\ \citenamefont {Murayama}}]{Kawasaki:2020qxm}%
  \BibitemOpen
  \bibfield  {author} {\bibinfo {author} {\bibfnamefont {M.}~\bibnamefont {Kawasaki}}, \bibinfo {author} {\bibfnamefont {K.}~\bibnamefont {Kohri}}, \bibinfo {author} {\bibfnamefont {T.}~\bibnamefont {Moroi}}, \bibinfo {author} {\bibfnamefont {K.}~\bibnamefont {Murai}}, \ and\ \bibinfo {author} {\bibfnamefont {H.}~\bibnamefont {Murayama}},\ }\href {\doibase 10.1088/1475-7516/2020/12/048} {\bibfield  {journal} {\bibinfo  {journal} {JCAP}\ }\textbf {\bibinfo {volume} {12}},\ \bibinfo {pages} {048} (\bibinfo {year} {2020})},\ \Eprint {http://arxiv.org/abs/2006.14803} {arXiv:2006.14803 [hep-ph]} \BibitemShut {NoStop}%
\bibitem [{\citenamefont {Pisanti}\ \emph {et~al.}(2008)\citenamefont {Pisanti}, \citenamefont {Cirillo}, \citenamefont {Esposito}, \citenamefont {Iocco}, \citenamefont {Mangano}, \citenamefont {Miele},\ and\ \citenamefont {Serpico}}]{Pisanti:2007hk}%
  \BibitemOpen
  \bibfield  {author} {\bibinfo {author} {\bibfnamefont {O.}~\bibnamefont {Pisanti}}, \bibinfo {author} {\bibfnamefont {A.}~\bibnamefont {Cirillo}}, \bibinfo {author} {\bibfnamefont {S.}~\bibnamefont {Esposito}}, \bibinfo {author} {\bibfnamefont {F.}~\bibnamefont {Iocco}}, \bibinfo {author} {\bibfnamefont {G.}~\bibnamefont {Mangano}}, \bibinfo {author} {\bibfnamefont {G.}~\bibnamefont {Miele}}, \ and\ \bibinfo {author} {\bibfnamefont {P.~D.}\ \bibnamefont {Serpico}},\ }\href {\doibase 10.1016/j.cpc.2008.02.015} {\bibfield  {journal} {\bibinfo  {journal} {Comput. Phys. Commun.}\ }\textbf {\bibinfo {volume} {178}},\ \bibinfo {pages} {956} (\bibinfo {year} {2008})},\ \Eprint {http://arxiv.org/abs/0705.0290} {arXiv:0705.0290 [astro-ph]} \BibitemShut {NoStop}%
\bibitem [{\citenamefont {Consiglio}\ \emph {et~al.}(2018)\citenamefont {Consiglio}, \citenamefont {de~Salas}, \citenamefont {Mangano}, \citenamefont {Miele}, \citenamefont {Pastor},\ and\ \citenamefont {Pisanti}}]{Consiglio:2017pot}%
  \BibitemOpen
  \bibfield  {author} {\bibinfo {author} {\bibfnamefont {R.}~\bibnamefont {Consiglio}}, \bibinfo {author} {\bibfnamefont {P.~F.}\ \bibnamefont {de~Salas}}, \bibinfo {author} {\bibfnamefont {G.}~\bibnamefont {Mangano}}, \bibinfo {author} {\bibfnamefont {G.}~\bibnamefont {Miele}}, \bibinfo {author} {\bibfnamefont {S.}~\bibnamefont {Pastor}}, \ and\ \bibinfo {author} {\bibfnamefont {O.}~\bibnamefont {Pisanti}},\ }\href {\doibase 10.1016/j.cpc.2018.06.022} {\bibfield  {journal} {\bibinfo  {journal} {Comput. Phys. Commun.}\ }\textbf {\bibinfo {volume} {233}},\ \bibinfo {pages} {237} (\bibinfo {year} {2018})},\ \Eprint {http://arxiv.org/abs/1712.04378} {arXiv:1712.04378 [astro-ph.CO]} \BibitemShut {NoStop}%
\bibitem [{\citenamefont {{Gariazzo}}\ \emph {et~al.}(2022)\citenamefont {{Gariazzo}}, \citenamefont {{F. de Salas}}, \citenamefont {{Pisanti}},\ and\ \citenamefont {{Consiglio}}}]{2022CoPhC.27108205G}%
  \BibitemOpen
  \bibfield  {author} {\bibinfo {author} {\bibfnamefont {S.}~\bibnamefont {{Gariazzo}}}, \bibinfo {author} {\bibfnamefont {P.}~\bibnamefont {{F. de Salas}}}, \bibinfo {author} {\bibfnamefont {O.}~\bibnamefont {{Pisanti}}}, \ and\ \bibinfo {author} {\bibfnamefont {R.}~\bibnamefont {{Consiglio}}},\ }\href {\doibase 10.1016/j.cpc.2021.108205} {\bibfield  {journal} {\bibinfo  {journal} {Computer Physics Communications}\ }\textbf {\bibinfo {volume} {271}},\ \bibinfo {eid} {108205} (\bibinfo {year} {2022})},\ \Eprint {http://arxiv.org/abs/2103.05027} {arXiv:2103.05027 [astro-ph.IM]} \BibitemShut {NoStop}%
\bibitem [{\citenamefont {Carr}\ \emph {et~al.}(2010)\citenamefont {Carr}, \citenamefont {Kohri}, \citenamefont {Sendouda},\ and\ \citenamefont {Yokoyama}}]{Carr:2009jm}%
  \BibitemOpen
  \bibfield  {author} {\bibinfo {author} {\bibfnamefont {B.~J.}\ \bibnamefont {Carr}}, \bibinfo {author} {\bibfnamefont {K.}~\bibnamefont {Kohri}}, \bibinfo {author} {\bibfnamefont {Y.}~\bibnamefont {Sendouda}}, \ and\ \bibinfo {author} {\bibfnamefont {J.}~\bibnamefont {Yokoyama}},\ }\href {\doibase 10.1103/PhysRevD.81.104019} {\bibfield  {journal} {\bibinfo  {journal} {Phys. Rev. D}\ }\textbf {\bibinfo {volume} {81}},\ \bibinfo {pages} {104019} (\bibinfo {year} {2010})},\ \Eprint {http://arxiv.org/abs/0912.5297} {arXiv:0912.5297 [astro-ph.CO]} \BibitemShut {NoStop}%
\bibitem [{\citenamefont {Carr}\ \emph {et~al.}(2021{\natexlab{b}})\citenamefont {Carr}, \citenamefont {Kohri}, \citenamefont {Sendouda},\ and\ \citenamefont {Yokoyama}}]{Carr:2020gox}%
  \BibitemOpen
  \bibfield  {author} {\bibinfo {author} {\bibfnamefont {B.}~\bibnamefont {Carr}}, \bibinfo {author} {\bibfnamefont {K.}~\bibnamefont {Kohri}}, \bibinfo {author} {\bibfnamefont {Y.}~\bibnamefont {Sendouda}}, \ and\ \bibinfo {author} {\bibfnamefont {J.}~\bibnamefont {Yokoyama}},\ }\href {\doibase 10.1088/1361-6633/ac1e31} {\bibfield  {journal} {\bibinfo  {journal} {Rept. Prog. Phys.}\ }\textbf {\bibinfo {volume} {84}},\ \bibinfo {pages} {116902} (\bibinfo {year} {2021}{\natexlab{b}})},\ \Eprint {http://arxiv.org/abs/2002.12778} {arXiv:2002.12778 [astro-ph.CO]} \BibitemShut {NoStop}%
\bibitem [{\citenamefont {Luo}\ \emph {et~al.}(2021)\citenamefont {Luo}, \citenamefont {Chen}, \citenamefont {Kusakabe},\ and\ \citenamefont {Kajino}}]{Luo:2020dlg}%
  \BibitemOpen
  \bibfield  {author} {\bibinfo {author} {\bibfnamefont {Y.}~\bibnamefont {Luo}}, \bibinfo {author} {\bibfnamefont {C.}~\bibnamefont {Chen}}, \bibinfo {author} {\bibfnamefont {M.}~\bibnamefont {Kusakabe}}, \ and\ \bibinfo {author} {\bibfnamefont {T.}~\bibnamefont {Kajino}},\ }\href {\doibase 10.1088/1475-7516/2021/05/042} {\bibfield  {journal} {\bibinfo  {journal} {JCAP}\ }\textbf {\bibinfo {volume} {05}},\ \bibinfo {pages} {042} (\bibinfo {year} {2021})},\ \Eprint {http://arxiv.org/abs/2011.10937} {arXiv:2011.10937 [astro-ph.CO]} \BibitemShut {NoStop}%
\bibitem [{\citenamefont {Mirbabayi}\ \emph {et~al.}(2020)\citenamefont {Mirbabayi}, \citenamefont {Gruzinov},\ and\ \citenamefont {Nore\~na}}]{Mirbabayi:2019uph}%
  \BibitemOpen
  \bibfield  {author} {\bibinfo {author} {\bibfnamefont {M.}~\bibnamefont {Mirbabayi}}, \bibinfo {author} {\bibfnamefont {A.}~\bibnamefont {Gruzinov}}, \ and\ \bibinfo {author} {\bibfnamefont {J.}~\bibnamefont {Nore\~na}},\ }\href {\doibase 10.1088/1475-7516/2020/03/017} {\bibfield  {journal} {\bibinfo  {journal} {JCAP}\ }\textbf {\bibinfo {volume} {03}},\ \bibinfo {pages} {017} (\bibinfo {year} {2020})},\ \Eprint {http://arxiv.org/abs/1901.05963} {arXiv:1901.05963 [astro-ph.CO]} \BibitemShut {NoStop}%
\bibitem [{\citenamefont {De~Luca}\ \emph {et~al.}(2019)\citenamefont {De~Luca}, \citenamefont {Desjacques}, \citenamefont {Franciolini}, \citenamefont {Malhotra},\ and\ \citenamefont {Riotto}}]{DeLuca:2019buf}%
  \BibitemOpen
  \bibfield  {author} {\bibinfo {author} {\bibfnamefont {V.}~\bibnamefont {De~Luca}}, \bibinfo {author} {\bibfnamefont {V.}~\bibnamefont {Desjacques}}, \bibinfo {author} {\bibfnamefont {G.}~\bibnamefont {Franciolini}}, \bibinfo {author} {\bibfnamefont {A.}~\bibnamefont {Malhotra}}, \ and\ \bibinfo {author} {\bibfnamefont {A.}~\bibnamefont {Riotto}},\ }\href {\doibase 10.1088/1475-7516/2019/05/018} {\bibfield  {journal} {\bibinfo  {journal} {JCAP}\ }\textbf {\bibinfo {volume} {05}},\ \bibinfo {pages} {018} (\bibinfo {year} {2019})},\ \Eprint {http://arxiv.org/abs/1903.01179} {arXiv:1903.01179 [astro-ph.CO]} \BibitemShut {NoStop}%
\bibitem [{\citenamefont {Page}\ and\ \citenamefont {Hawking}(1976)}]{Page:1976wx}%
  \BibitemOpen
  \bibfield  {author} {\bibinfo {author} {\bibfnamefont {D.~N.}\ \bibnamefont {Page}}\ and\ \bibinfo {author} {\bibfnamefont {S.~W.}\ \bibnamefont {Hawking}},\ }\href {\doibase 10.1086/154350} {\bibfield  {journal} {\bibinfo  {journal} {Astrophys. J.}\ }\textbf {\bibinfo {volume} {206}},\ \bibinfo {pages} {1} (\bibinfo {year} {1976})}\BibitemShut {NoStop}%
\bibitem [{\citenamefont {Mazde}\ and\ \citenamefont {Visinelli}(2023)}]{Mazde:2022sdx}%
  \BibitemOpen
  \bibfield  {author} {\bibinfo {author} {\bibfnamefont {K.}~\bibnamefont {Mazde}}\ and\ \bibinfo {author} {\bibfnamefont {L.}~\bibnamefont {Visinelli}},\ }\href {\doibase 10.1088/1475-7516/2023/01/021} {\bibfield  {journal} {\bibinfo  {journal} {JCAP}\ }\textbf {\bibinfo {volume} {01}},\ \bibinfo {pages} {021} (\bibinfo {year} {2023})},\ \Eprint {http://arxiv.org/abs/2209.14307} {arXiv:2209.14307 [astro-ph.CO]} \BibitemShut {NoStop}%
\bibitem [{\citenamefont {Page}(1977)}]{Page:1977um}%
  \BibitemOpen
  \bibfield  {author} {\bibinfo {author} {\bibfnamefont {D.~N.}\ \bibnamefont {Page}},\ }\href {\doibase 10.1103/PhysRevD.16.2402} {\bibfield  {journal} {\bibinfo  {journal} {Phys. Rev. D}\ }\textbf {\bibinfo {volume} {16}},\ \bibinfo {pages} {2402} (\bibinfo {year} {1977})}\BibitemShut {NoStop}%
\bibitem [{\citenamefont {Aghanim}\ \emph {et~al.}(2020)\citenamefont {Aghanim} \emph {et~al.}}]{Planck:2018vyg}%
  \BibitemOpen
  \bibfield  {author} {\bibinfo {author} {\bibfnamefont {N.}~\bibnamefont {Aghanim}} \emph {et~al.} (\bibinfo {collaboration} {Planck}),\ }\href {\doibase 10.1051/0004-6361/201833910} {\bibfield  {journal} {\bibinfo  {journal} {Astron. Astrophys.}\ }\textbf {\bibinfo {volume} {641}},\ \bibinfo {pages} {A6} (\bibinfo {year} {2020})},\ \bibinfo {note} {[Erratum: Astron.Astrophys. 652, C4 (2021)]},\ \Eprint {http://arxiv.org/abs/1807.06209} {arXiv:1807.06209 [astro-ph.CO]} \BibitemShut {NoStop}%
\bibitem [{\citenamefont {Iocco}\ \emph {et~al.}(2007)\citenamefont {Iocco}, \citenamefont {Mangano}, \citenamefont {Miele}, \citenamefont {Pisanti},\ and\ \citenamefont {Serpico}}]{Iocco:2007km}%
  \BibitemOpen
  \bibfield  {author} {\bibinfo {author} {\bibfnamefont {F.}~\bibnamefont {Iocco}}, \bibinfo {author} {\bibfnamefont {G.}~\bibnamefont {Mangano}}, \bibinfo {author} {\bibfnamefont {G.}~\bibnamefont {Miele}}, \bibinfo {author} {\bibfnamefont {O.}~\bibnamefont {Pisanti}}, \ and\ \bibinfo {author} {\bibfnamefont {P.~D.}\ \bibnamefont {Serpico}},\ }\href {\doibase 10.1103/PhysRevD.75.087304} {\bibfield  {journal} {\bibinfo  {journal} {Phys. Rev. D}\ }\textbf {\bibinfo {volume} {75}},\ \bibinfo {pages} {087304} (\bibinfo {year} {2007})},\ \Eprint {http://arxiv.org/abs/astro-ph/0702090} {arXiv:astro-ph/0702090} \BibitemShut {NoStop}%
\bibitem [{\citenamefont {Schramm}\ and\ \citenamefont {Turner}(1998)}]{Schramm:1997vs}%
  \BibitemOpen
  \bibfield  {author} {\bibinfo {author} {\bibfnamefont {D.~N.}\ \bibnamefont {Schramm}}\ and\ \bibinfo {author} {\bibfnamefont {M.~S.}\ \bibnamefont {Turner}},\ }\href {\doibase 10.1103/RevModPhys.70.303} {\bibfield  {journal} {\bibinfo  {journal} {Rev. Mod. Phys.}\ }\textbf {\bibinfo {volume} {70}},\ \bibinfo {pages} {303} (\bibinfo {year} {1998})},\ \Eprint {http://arxiv.org/abs/astro-ph/9706069} {arXiv:astro-ph/9706069} \BibitemShut {NoStop}%
\bibitem [{\citenamefont {Steigman}(2007)}]{Steigman:2007xt}%
  \BibitemOpen
  \bibfield  {author} {\bibinfo {author} {\bibfnamefont {G.}~\bibnamefont {Steigman}},\ }\href {\doibase 10.1146/annurev.nucl.56.080805.140437} {\bibfield  {journal} {\bibinfo  {journal} {Ann. Rev. Nucl. Part. Sci.}\ }\textbf {\bibinfo {volume} {57}},\ \bibinfo {pages} {463} (\bibinfo {year} {2007})},\ \Eprint {http://arxiv.org/abs/0712.1100} {arXiv:0712.1100 [astro-ph]} \BibitemShut {NoStop}%
\bibitem [{\citenamefont {Iocco}\ \emph {et~al.}(2009)\citenamefont {Iocco}, \citenamefont {Mangano}, \citenamefont {Miele}, \citenamefont {Pisanti},\ and\ \citenamefont {Serpico}}]{Iocco:2008va}%
  \BibitemOpen
  \bibfield  {author} {\bibinfo {author} {\bibfnamefont {F.}~\bibnamefont {Iocco}}, \bibinfo {author} {\bibfnamefont {G.}~\bibnamefont {Mangano}}, \bibinfo {author} {\bibfnamefont {G.}~\bibnamefont {Miele}}, \bibinfo {author} {\bibfnamefont {O.}~\bibnamefont {Pisanti}}, \ and\ \bibinfo {author} {\bibfnamefont {P.~D.}\ \bibnamefont {Serpico}},\ }\href {\doibase 10.1016/j.physrep.2009.02.002} {\bibfield  {journal} {\bibinfo  {journal} {Phys. Rept.}\ }\textbf {\bibinfo {volume} {472}},\ \bibinfo {pages} {1} (\bibinfo {year} {2009})},\ \Eprint {http://arxiv.org/abs/0809.0631} {arXiv:0809.0631 [astro-ph]} \BibitemShut {NoStop}%
\bibitem [{\citenamefont {Cyburt}\ \emph {et~al.}(2016)\citenamefont {Cyburt}, \citenamefont {Fields}, \citenamefont {Olive},\ and\ \citenamefont {Yeh}}]{Cyburt:2015mya}%
  \BibitemOpen
  \bibfield  {author} {\bibinfo {author} {\bibfnamefont {R.~H.}\ \bibnamefont {Cyburt}}, \bibinfo {author} {\bibfnamefont {B.~D.}\ \bibnamefont {Fields}}, \bibinfo {author} {\bibfnamefont {K.~A.}\ \bibnamefont {Olive}}, \ and\ \bibinfo {author} {\bibfnamefont {T.-H.}\ \bibnamefont {Yeh}},\ }\href {\doibase 10.1103/RevModPhys.88.015004} {\bibfield  {journal} {\bibinfo  {journal} {Rev. Mod. Phys.}\ }\textbf {\bibinfo {volume} {88}},\ \bibinfo {pages} {015004} (\bibinfo {year} {2016})},\ \Eprint {http://arxiv.org/abs/1505.01076} {arXiv:1505.01076 [astro-ph.CO]} \BibitemShut {NoStop}%
\bibitem [{\citenamefont {Pospelov}\ and\ \citenamefont {Pradler}(2010)}]{Pospelov:2010hj}%
  \BibitemOpen
  \bibfield  {author} {\bibinfo {author} {\bibfnamefont {M.}~\bibnamefont {Pospelov}}\ and\ \bibinfo {author} {\bibfnamefont {J.}~\bibnamefont {Pradler}},\ }\href {\doibase 10.1146/annurev.nucl.012809.104521} {\bibfield  {journal} {\bibinfo  {journal} {Ann. Rev. Nucl. Part. Sci.}\ }\textbf {\bibinfo {volume} {60}},\ \bibinfo {pages} {539} (\bibinfo {year} {2010})},\ \Eprint {http://arxiv.org/abs/1011.1054} {arXiv:1011.1054 [hep-ph]} \BibitemShut {NoStop}%
\bibitem [{\citenamefont {Workman}\ \emph {et~al.}(2022)\citenamefont {Workman} \emph {et~al.}}]{ParticleDataGroup:2022pth}%
  \BibitemOpen
  \bibfield  {author} {\bibinfo {author} {\bibfnamefont {R.~L.}\ \bibnamefont {Workman}} \emph {et~al.} (\bibinfo {collaboration} {Particle Data Group}),\ }\href {\doibase 10.1093/ptep/ptac097} {\bibfield  {journal} {\bibinfo  {journal} {PTEP}\ }\textbf {\bibinfo {volume} {2022}},\ \bibinfo {pages} {083C01} (\bibinfo {year} {2022})}\BibitemShut {NoStop}%
\bibitem [{\citenamefont {Esmailzadeh}\ \emph {et~al.}(1991)\citenamefont {Esmailzadeh}, \citenamefont {Starkman},\ and\ \citenamefont {Dimopoulos}}]{Esmailzadeh:1990hf}%
  \BibitemOpen
  \bibfield  {author} {\bibinfo {author} {\bibfnamefont {R.}~\bibnamefont {Esmailzadeh}}, \bibinfo {author} {\bibfnamefont {G.~D.}\ \bibnamefont {Starkman}}, \ and\ \bibinfo {author} {\bibfnamefont {S.}~\bibnamefont {Dimopoulos}},\ }\href@noop {} {\bibfield  {journal} {\bibinfo  {journal} {Astrophys. J.}\ }\textbf {\bibinfo {volume} {378}},\ \bibinfo {pages} {504} (\bibinfo {year} {1991})}\BibitemShut {NoStop}%
\bibitem [{\citenamefont {Mukhanov}(2004)}]{Mukhanov:2003xs}%
  \BibitemOpen
  \bibfield  {author} {\bibinfo {author} {\bibfnamefont {V.~F.}\ \bibnamefont {Mukhanov}},\ }\href {\doibase 10.1023/B:IJTP.0000048169.69609.77} {\bibfield  {journal} {\bibinfo  {journal} {Int. J. Theor. Phys.}\ }\textbf {\bibinfo {volume} {43}},\ \bibinfo {pages} {669} (\bibinfo {year} {2004})},\ \Eprint {http://arxiv.org/abs/astro-ph/0303073} {arXiv:astro-ph/0303073} \BibitemShut {NoStop}%
\bibitem [{\citenamefont {{Adams}}(1976)}]{1976A&A....50..461A}%
  \BibitemOpen
  \bibfield  {author} {\bibinfo {author} {\bibfnamefont {T.~F.}\ \bibnamefont {{Adams}}},\ }\href@noop {} {\bibfield  {journal} {\bibinfo  {journal} {Astron. Astrophys.}\ }\textbf {\bibinfo {volume} {50}},\ \bibinfo {pages} {461} (\bibinfo {year} {1976})}\BibitemShut {NoStop}%
\bibitem [{\citenamefont {Serpico}\ \emph {et~al.}(2004)\citenamefont {Serpico}, \citenamefont {Esposito}, \citenamefont {Iocco}, \citenamefont {Mangano}, \citenamefont {Miele},\ and\ \citenamefont {Pisanti}}]{Serpico:2004gx}%
  \BibitemOpen
  \bibfield  {author} {\bibinfo {author} {\bibfnamefont {P.~D.}\ \bibnamefont {Serpico}}, \bibinfo {author} {\bibfnamefont {S.}~\bibnamefont {Esposito}}, \bibinfo {author} {\bibfnamefont {F.}~\bibnamefont {Iocco}}, \bibinfo {author} {\bibfnamefont {G.}~\bibnamefont {Mangano}}, \bibinfo {author} {\bibfnamefont {G.}~\bibnamefont {Miele}}, \ and\ \bibinfo {author} {\bibfnamefont {O.}~\bibnamefont {Pisanti}},\ }\href {\doibase 10.1088/1475-7516/2004/12/010} {\bibfield  {journal} {\bibinfo  {journal} {JCAP}\ }\textbf {\bibinfo {volume} {12}},\ \bibinfo {pages} {010} (\bibinfo {year} {2004})},\ \Eprint {http://arxiv.org/abs/astro-ph/0408076} {arXiv:astro-ph/0408076} \BibitemShut {NoStop}%
\bibitem [{\citenamefont {Masina}(2020)}]{Masina:2020xhk}%
  \BibitemOpen
  \bibfield  {author} {\bibinfo {author} {\bibfnamefont {I.}~\bibnamefont {Masina}},\ }\href {\doibase 10.1140/epjp/s13360-020-00564-9} {\bibfield  {journal} {\bibinfo  {journal} {Eur. Phys. J. Plus}\ }\textbf {\bibinfo {volume} {135}},\ \bibinfo {pages} {552} (\bibinfo {year} {2020})},\ \Eprint {http://arxiv.org/abs/2004.04740} {arXiv:2004.04740 [hep-ph]} \BibitemShut {NoStop}%
\bibitem [{\citenamefont {Iocco}(2012)}]{Iocco:2012vg}%
  \BibitemOpen
  \bibfield  {author} {\bibinfo {author} {\bibfnamefont {F.}~\bibnamefont {Iocco}},\ }\href@noop {} {\bibfield  {journal} {\bibinfo  {journal} {Mem. Soc. Astron. Ital. Suppl.}\ }\textbf {\bibinfo {volume} {22}},\ \bibinfo {pages} {19} (\bibinfo {year} {2012})},\ \Eprint {http://arxiv.org/abs/1206.2396} {arXiv:1206.2396 [astro-ph.GA]} \BibitemShut {NoStop}%
\bibitem [{\citenamefont {Zyla}\ \emph {et~al.}(2020)\citenamefont {Zyla} \emph {et~al.}}]{10.1093/ptep/ptaa104}%
  \BibitemOpen
  \bibfield  {author} {\bibinfo {author} {\bibfnamefont {P.~A.}\ \bibnamefont {Zyla}} \emph {et~al.} (\bibinfo {collaboration} {Particle Data Group}),\ }\href {\doibase 10.1093/ptep/ptaa104} {\bibfield  {journal} {\bibinfo  {journal} {Progress of Theoretical and Experimental Physics}\ }\textbf {\bibinfo {volume} {2020}},\ \bibinfo {pages} {083C01} (\bibinfo {year} {2020})},\ \Eprint {http://arxiv.org/abs/https://academic.oup.com/ptep/article-pdf/2020/8/083C01/34673722/ptaa104.pdf} {https://academic.oup.com/ptep/article-pdf/2020/8/083C01/34673722/ptaa104.pdf} \BibitemShut {NoStop}%
\bibitem [{\citenamefont {Dom\`enech}\ \emph {et~al.}(2021)\citenamefont {Dom\`enech}, \citenamefont {Takhistov},\ and\ \citenamefont {Sasaki}}]{Domenech:2021wkk}%
  \BibitemOpen
  \bibfield  {author} {\bibinfo {author} {\bibfnamefont {G.}~\bibnamefont {Dom\`enech}}, \bibinfo {author} {\bibfnamefont {V.}~\bibnamefont {Takhistov}}, \ and\ \bibinfo {author} {\bibfnamefont {M.}~\bibnamefont {Sasaki}},\ }\href {\doibase 10.1016/j.physletb.2021.136722} {\bibfield  {journal} {\bibinfo  {journal} {Phys. Lett. B}\ }\textbf {\bibinfo {volume} {823}},\ \bibinfo {pages} {136722} (\bibinfo {year} {2021})},\ \Eprint {http://arxiv.org/abs/2105.06816} {arXiv:2105.06816 [astro-ph.CO]} \BibitemShut {NoStop}%
\bibitem [{\citenamefont {Kohri}\ and\ \citenamefont {Yokoyama}(2000)}]{Kohri:1999ex}%
  \BibitemOpen
  \bibfield  {author} {\bibinfo {author} {\bibfnamefont {K.}~\bibnamefont {Kohri}}\ and\ \bibinfo {author} {\bibfnamefont {J.}~\bibnamefont {Yokoyama}},\ }\href {\doibase 10.1103/PhysRevD.61.023501} {\bibfield  {journal} {\bibinfo  {journal} {Phys. Rev. D}\ }\textbf {\bibinfo {volume} {61}},\ \bibinfo {pages} {023501} (\bibinfo {year} {2000})},\ \Eprint {http://arxiv.org/abs/astro-ph/9908160} {arXiv:astro-ph/9908160} \BibitemShut {NoStop}%
\bibitem [{\citenamefont {{Vainer}}\ and\ \citenamefont {{Naselskii}}(1978)}]{1978SvA....22..138V}%
  \BibitemOpen
  \bibfield  {author} {\bibinfo {author} {\bibfnamefont {B.~V.}\ \bibnamefont {{Vainer}}}\ and\ \bibinfo {author} {\bibfnamefont {P.~D.}\ \bibnamefont {{Naselskii}}},\ }\href@noop {} {\bibfield  {journal} {\bibinfo  {journal} {Soviet Astron.}\ }\textbf {\bibinfo {volume} {22}},\ \bibinfo {pages} {138} (\bibinfo {year} {1978})}\BibitemShut {NoStop}%
\bibitem [{\citenamefont {Mastrototaro}\ \emph {et~al.}(2021)\citenamefont {Mastrototaro}, \citenamefont {Serpico}, \citenamefont {Mirizzi},\ and\ \citenamefont {Saviano}}]{Mastrototaro:2021wzl}%
  \BibitemOpen
  \bibfield  {author} {\bibinfo {author} {\bibfnamefont {L.}~\bibnamefont {Mastrototaro}}, \bibinfo {author} {\bibfnamefont {P.~D.}\ \bibnamefont {Serpico}}, \bibinfo {author} {\bibfnamefont {A.}~\bibnamefont {Mirizzi}}, \ and\ \bibinfo {author} {\bibfnamefont {N.}~\bibnamefont {Saviano}},\ }\href {\doibase 10.1103/PhysRevD.104.016026} {\bibfield  {journal} {\bibinfo  {journal} {Phys. Rev. D}\ }\textbf {\bibinfo {volume} {104}},\ \bibinfo {pages} {016026} (\bibinfo {year} {2021})},\ \Eprint {http://arxiv.org/abs/2104.11752} {arXiv:2104.11752 [hep-ph]} \BibitemShut {NoStop}%
\bibitem [{\citenamefont {Kawasaki}\ \emph {et~al.}(1999)\citenamefont {Kawasaki}, \citenamefont {Kohri},\ and\ \citenamefont {Sugiyama}}]{Kawasaki:1999na}%
  \BibitemOpen
  \bibfield  {author} {\bibinfo {author} {\bibfnamefont {M.}~\bibnamefont {Kawasaki}}, \bibinfo {author} {\bibfnamefont {K.}~\bibnamefont {Kohri}}, \ and\ \bibinfo {author} {\bibfnamefont {N.}~\bibnamefont {Sugiyama}},\ }\href {\doibase 10.1103/PhysRevLett.82.4168} {\bibfield  {journal} {\bibinfo  {journal} {Phys. Rev. Lett.}\ }\textbf {\bibinfo {volume} {82}},\ \bibinfo {pages} {4168} (\bibinfo {year} {1999})},\ \Eprint {http://arxiv.org/abs/astro-ph/9811437} {arXiv:astro-ph/9811437} \BibitemShut {NoStop}%
\bibitem [{\citenamefont {Kawasaki}\ \emph {et~al.}(2000)\citenamefont {Kawasaki}, \citenamefont {Kohri},\ and\ \citenamefont {Sugiyama}}]{Kawasaki:2000en}%
  \BibitemOpen
  \bibfield  {author} {\bibinfo {author} {\bibfnamefont {M.}~\bibnamefont {Kawasaki}}, \bibinfo {author} {\bibfnamefont {K.}~\bibnamefont {Kohri}}, \ and\ \bibinfo {author} {\bibfnamefont {N.}~\bibnamefont {Sugiyama}},\ }\href {\doibase 10.1103/PhysRevD.62.023506} {\bibfield  {journal} {\bibinfo  {journal} {Phys. Rev. D}\ }\textbf {\bibinfo {volume} {62}},\ \bibinfo {pages} {023506} (\bibinfo {year} {2000})},\ \Eprint {http://arxiv.org/abs/astro-ph/0002127} {arXiv:astro-ph/0002127} \BibitemShut {NoStop}%
\bibitem [{\citenamefont {de~Salas}\ \emph {et~al.}(2015)\citenamefont {de~Salas}, \citenamefont {Lattanzi}, \citenamefont {Mangano}, \citenamefont {Miele}, \citenamefont {Pastor},\ and\ \citenamefont {Pisanti}}]{deSalas:2015glj}%
  \BibitemOpen
  \bibfield  {author} {\bibinfo {author} {\bibfnamefont {P.~F.}\ \bibnamefont {de~Salas}}, \bibinfo {author} {\bibfnamefont {M.}~\bibnamefont {Lattanzi}}, \bibinfo {author} {\bibfnamefont {G.}~\bibnamefont {Mangano}}, \bibinfo {author} {\bibfnamefont {G.}~\bibnamefont {Miele}}, \bibinfo {author} {\bibfnamefont {S.}~\bibnamefont {Pastor}}, \ and\ \bibinfo {author} {\bibfnamefont {O.}~\bibnamefont {Pisanti}},\ }\href {\doibase 10.1103/PhysRevD.92.123534} {\bibfield  {journal} {\bibinfo  {journal} {Phys. Rev. D}\ }\textbf {\bibinfo {volume} {92}},\ \bibinfo {pages} {123534} (\bibinfo {year} {2015})},\ \Eprint {http://arxiv.org/abs/1511.00672} {arXiv:1511.00672 [astro-ph.CO]} \BibitemShut {NoStop}%
\bibitem [{\citenamefont {Hasegawa}\ \emph {et~al.}(2019)\citenamefont {Hasegawa}, \citenamefont {Hiroshima}, \citenamefont {Kohri}, \citenamefont {Hansen}, \citenamefont {Tram},\ and\ \citenamefont {Hannestad}}]{Hasegawa:2019jsa}%
  \BibitemOpen
  \bibfield  {author} {\bibinfo {author} {\bibfnamefont {T.}~\bibnamefont {Hasegawa}}, \bibinfo {author} {\bibfnamefont {N.}~\bibnamefont {Hiroshima}}, \bibinfo {author} {\bibfnamefont {K.}~\bibnamefont {Kohri}}, \bibinfo {author} {\bibfnamefont {R.~S.~L.}\ \bibnamefont {Hansen}}, \bibinfo {author} {\bibfnamefont {T.}~\bibnamefont {Tram}}, \ and\ \bibinfo {author} {\bibfnamefont {S.}~\bibnamefont {Hannestad}},\ }\href {\doibase 10.1088/1475-7516/2019/12/012} {\bibfield  {journal} {\bibinfo  {journal} {JCAP}\ }\textbf {\bibinfo {volume} {12}},\ \bibinfo {pages} {012} (\bibinfo {year} {2019})},\ \Eprint {http://arxiv.org/abs/1908.10189} {arXiv:1908.10189 [hep-ph]} \BibitemShut {NoStop}%
\bibitem [{\citenamefont {Niemeyer}\ and\ \citenamefont {Jedamzik}(1998)}]{Niemeyer:1997mt}%
  \BibitemOpen
  \bibfield  {author} {\bibinfo {author} {\bibfnamefont {J.~C.}\ \bibnamefont {Niemeyer}}\ and\ \bibinfo {author} {\bibfnamefont {K.}~\bibnamefont {Jedamzik}},\ }\href {\doibase 10.1103/PhysRevLett.80.5481} {\bibfield  {journal} {\bibinfo  {journal} {Phys. Rev. Lett.}\ }\textbf {\bibinfo {volume} {80}},\ \bibinfo {pages} {5481} (\bibinfo {year} {1998})},\ \Eprint {http://arxiv.org/abs/astro-ph/9709072} {arXiv:astro-ph/9709072} \BibitemShut {NoStop}%
\end{thebibliography}%

\end{document}